\documentclass[aps,pra,twocolumn,floats,showpacs]{revtex4}

\usepackage{graphicx}
\usepackage{amsmath}
\usepackage{amssymb}
\usepackage{color}
\usepackage{ulem}
\DeclareMathAlphabet{\mathantt}{OT1}{antt}{li}{it}

\newcommand{\ve}{{\varepsilon}}
\renewcommand{\a}{{\alpha}}

\newcommand{\w}{{\omega}}

\newcommand{\g}{{\gamma}}

\newcommand{\W}{{\Omega}}
\renewcommand{\d}{{\delta}}

\def\K{{\mathfrak K}}

\newcommand{\cmm}{{cm$^{-1}$}}
\newcommand{\mum}{{$\mu$m}}

\begin{document}

\title{Photonic crystal optics in cold atomic gases}
\author{Marina Litinskaya, Evgeny A. Shapiro}
\affiliation{Department of Chemistry, University of British Columbia, Vancouver, V6T 1Z1, Canada}

\begin{abstract}
We describe propagation of light in a gas with periodic density
modulation, demonstrating photonic-crystal-like refraction with
negative refraction angles. We address the role of poorly defined
boundaries and damping, and derive an optical analog of the
quantum adiabatic theorem. For Cs atoms in an optical lattice, we
show that relying on semi-adiabatic propagation one can excite and
spatially split positively and negatively refracting modes at
experimentally available gas densities.

\end{abstract}


\maketitle

{\textbf{\textit{ Introduction.}} A wave is negatively refracted
at an interface of two media if its group velocity component along
the interface changes its sign  \cite{NR}, see Fig.
\ref{f-Cheburashki}(a). Fascinating optical effects based on
negative refraction (NR) include invisibility \cite{invisibility},
near-field focusing with planar devices \cite{lens}, seeing around
a corner \cite{corner}, and superprism \cite{superprism}. This
work is aimed at achieving similar effects in a gas. Using laser
fields instead of nanofabrication for preparing the sample will
enable dynamical real-time control at a distance in the optical
frequency domain \cite{Turki,Chirality}. New applications such as
nonlinear spectroscopy with backward propagating signal
\cite{Kravtsov} may become available.

Photonic-crystal (PC) metamaterials offer a route to NR
\cite{Foteinopoulou}. Both negatively and positively refracted
modes appear due to periodic modulation of the dielectric
constant, see Fig.\ref{f-Cheburashki}(b). We shall call them
``N-'' and ``P-''modes, respectively, implying both the
negative-like refraction in 1D PCs and true NR in 2D, where
Snell's law with a negative refractive index is in effect
\cite{Notomi}.
Below we theoretically demonstrate the possibility of
negative-like refraction in a cold gas trapped in an optical
lattice. 
Photonic band gap is routinely observed in such systems
\cite{EIT-exp}.  We study NR in the proof of principle case of 1D
periodicity; implementation in 2D and 3D is straightforward.  The
scheme offers relatively low gas densities, simple design, and a
large frequency window of negative refraction.

Implementation in a gas brings two challenges absent in solid
materials. First, due to low densities, a significant modulation
of the dielectric constant is only possible in a narrow vicinity
of a resonance, where absorption is high. This resonant absorption
can be overcome using schemes based on Electromagnetically Induced
Transparency (EIT) \cite{Lukin-LaRocca}.
Here we study another route: For a thermal gas trapped in a 1D
optical lattice, we consider relatively large detuning from the
resonance so that both absorption and modulation of the refractive
index are moderate. The negative-like refraction emerges due to
periodicity of the lattice potential. In this arrangement, the
required gas density turns out to be experimentally achievable
$10^{13}~{\rm cm}^{-3}$ \cite{10^13}, three to five orders of
magnitude lower than in the earlier proposals \cite{Turki,
Chirality}. As discussed below, the advantage comes both from not
relying on the weak magnetic response of the gas, and from the
N-mode being immune to resonant absorption.

The second challenge, common for all gaseous samples and never
treated before, is due to poorly defined boundaries in a gas
cloud. If light penetrates the cloud adiabatically, then only
P-mode is excited and negative refraction is never achieved.
Below we derive an analog of the quantum adiabatic theorem for coupled propagation of N- and P-modes, and study the 
dynamics of energy transfer between them. We show that with
experimentally achievable conditions one can realize propagation
possessing both adiabatic and non-adiabatic features, thus
providing transfer into the N-mode, and simultaneously avoiding
unwanted reflection at boundaries.


\textbf{\textit{Waves in resonant periodic arrays.}}
A 1D PC with a period $a$ comparable with a half of the light
wavelength $\lambda$ (Fig. \ref{f-Cheburashki}(b)) acts as a
volume diffraction grating and supports P- and N-diffracted modes
\cite{Russel, Notomi}. We model an infinite (so that the  boundary
is irrelevant) grating as a periodic set of delta-like
perturbations in the dielectric constant:

%
\begin{equation} \label{eps(z,w)}
\varepsilon(z; \omega) = 1 + \varepsilon_{c} {d} \sum\limits_{n}
\delta(z - n a) = 1 + \frac{d}{a}\varepsilon_{c} \sum_n e^{inz
{G}}.
\end{equation}
Here $G = 2\pi/a$, and $\varepsilon_{c}$ is the dielectric
contrast between the layers of the grating and vacuum. Applying
modal approach \cite{Chu-Tamir}, we find an analytical solution
for the Maxwell equations for TE-polarized waves, ${\bf E} = (0,
E, 0)$, ${\bf H} = (H_x, 0, H_z)$, as follows. Due to periodicity
along $z$, the eigenstates are Floquet-Bloch (FB) modes
\cite{Russel} consisting of partial waves (marked by index $m$) of
the form
\begin{equation}\label{E-field}
E(z; t) = \sum\limits_m C_m\ \exp\biggl[i(k_x x + (k_z + m G)z - \omega t)\biggr],
\end{equation}
with the amplitudes $C_m$ obtained by substituting (\ref{eps(z,w)}) and
(\ref{E-field}) into the Maxwell wave equation:
\begin{equation}\label{C_m}
C_m = \frac{\varepsilon_c \omega^2\ \bigl( \sum_lC_l \bigr) }{c^2[
k_x^2 + (k_z + mG)^2] - \omega^2}~.
\end{equation}
The values $\omega$, $k_x$ and $k_z$ satisfy the dispersion
equation, obtained by summing (\ref{C_m}) over $m$ and cancelling
$\sum_m C_m$ and $\sum_l C_l$. The remaining sum over $m$ can be
calculated analytically. We write the result as:
\begin{equation}\label{dispersion}
\cos ak_z = \cos ak_{z}^{(vac)} -
\frac{d}{a}\frac{\varepsilon_c}{2} \left( \frac{\omega a}{c}
\right)^2 \frac{\sin a k_{z}^{(vac)}}{a k_{z}^{(vac)}}~,
\end{equation}
where $k_{z}^{(vac)}=\sqrt{\frac{\omega^2}{c^2} - k_x^2}$. FB
modes of the form (\ref{E-field}-\ref{C_m}) are sketched in
Fig.~\ref{f-Cheburashki}(b). The arrows show phase velocity
directions.
 The group velocity is the same for all partial waves in a mode.
Amplitude of $m$-th partial wave is determined by the
corresponding denominator in RHS of Eq.(\ref{C_m}), i.e. by how
close to $\omega^2/c^2$ each $k_x^2 + (k_z+mG)^2$ is, and is shown
schematically by the arrow thickness.

The number of different FB modes with the same value of $k_z$ and their propagation directions can be deduced from the
 equi-frequency surface (EFS) for Eq.(\ref{dispersion})
in the $k_x,k_z$ plane \cite{Russel, Notomi, Chu-Tamir}.
Fig.~\ref{f-Cheburashki}(c) shows an example EFS for  the
frequency $\omega = 1.25 \pi c/a$ and strong modulation of
$\varepsilon$. The periodicity results in the EFS diagram with the
property $\omega(k_x, k_z) = \omega(k_x, k_z+mG)$ for all $m$:
instead of a single circle $k_x^2+k_z^2=\omega^2/c^2$ (green
dashed circle in the figure), EFS consists of a series of circles
corresponding to different partial waves. In addition, the
modulation opens Bragg gaps at $k_{z} = G/2 + m G$, which change
the topology of the EFS: Instead of intersecting circles
corresponding to the bare photon dispersion repeated along $k_z$,
there is a series of smaller inner ellipses embraced by merging
outer parts of the circles.
Each FB mode of the PC is characterized by a specific value of
$k_x$, and a set of $k_z = k_{z0} + mG$, $m=0, \pm 1, \pm 2, ...$.
In Fig.~\ref{f-Cheburashki}(c) they are represented by points on
the EFS. Points C and C$'$ belong to the same mode: C marks
its partial wave with $m=0$, C$'$ marks $m=-1$.

Consider a light beam incident from $x = -\infty$ at an angle
$\phi$ as shown in Fig.\ref{f-Cheburashki}(b). This beam is
represented by point A on the green dashed circle $\sqrt{k_{x
0}^2+k_{z0}^2}=\omega/c$ in Fig.\ref{f-Cheburashki}(c). At the
boundary, it can couple to all the FB modes with the same value of
$k_{z0} = (\omega/c) \sin \phi$.  Intersections of the line $k_z =
k_{z0}$ (dotted blue line) with the EFS determine the $k_x$ values
of the eigenmodes coupled to the probe. Normals to the EFS
(depictes by arrows in Fig.~\ref{f-Cheburashki}(c)) determine the
group velocity directions of these modes \cite{Russel, Notomi,
Chu-Tamir}. Figure shows that for $k_{z0}$ close to $\pi/a$ there
are two normal modes with positive $x$-component of the group
velocity: an N-mode (with $k_{x} \equiv k_{N}$, point C) and a
P-mode (with $k_{x} \equiv k_{P}$, point B).  Note that the N-mode
exhibits negative-like refraction. Points $D$ and $E$ with the
same $k_{z0}$ and $k_x=-k_{N,P}$ correspond to two reflected
waves, marked as ${\rm N_{ref}}$ and ${\rm P_{ref}}$.

%
\begin{figure}
\centering
\includegraphics[width=0.85\columnwidth]{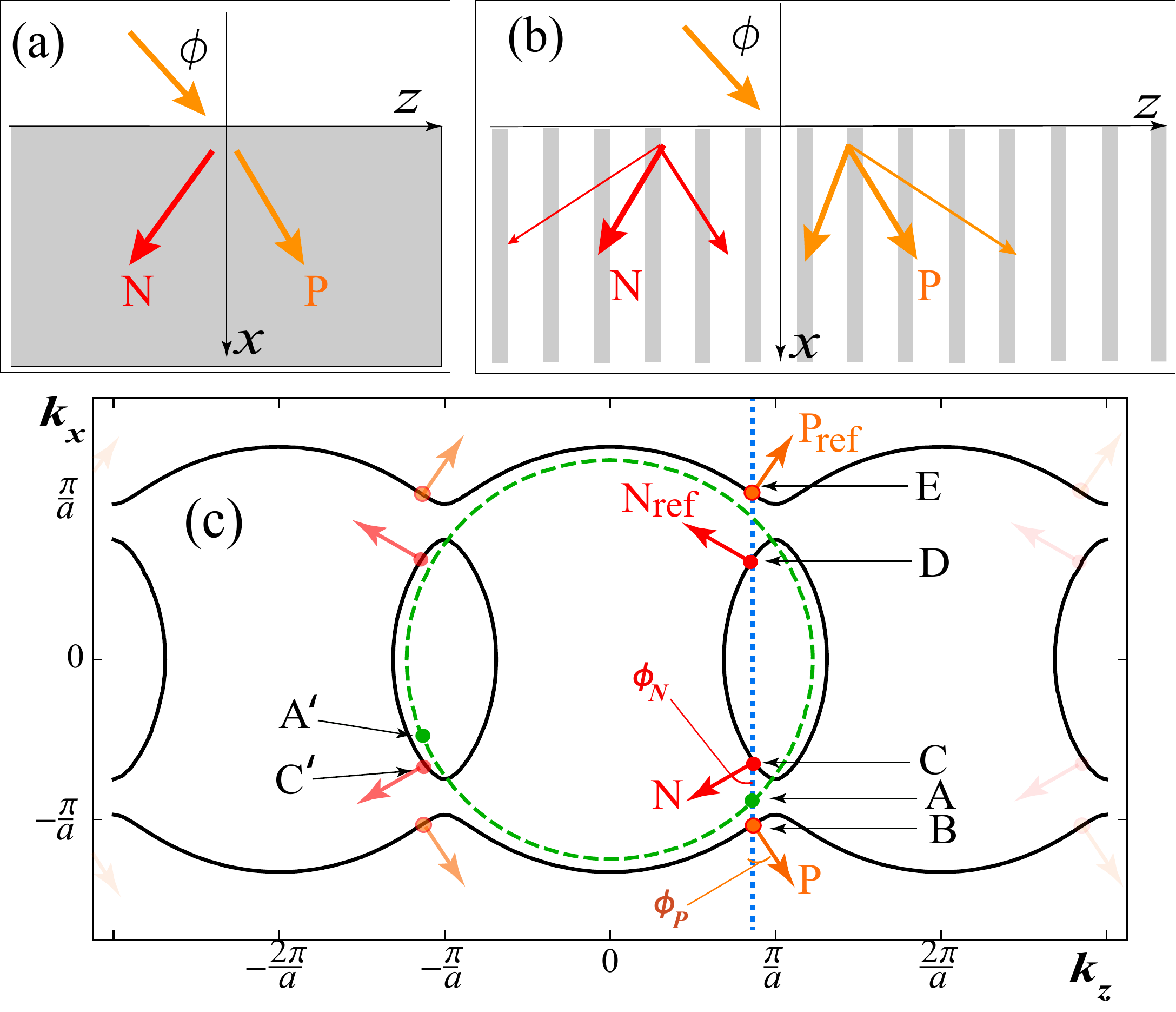}

\caption{(Color online) (a) Negative (red) and positive (gold) refraction at an
interface. (b) Negative and positive Floquet-Bloch
modes in a 1D photonic crystal. (c) Solid black line: EFS in a 1D
PC. Green dashed circle: EFS for light in vacuum. Red and gold
arrows show the propagation directions of the N- and P-modes.
Points A, A$^{'}$ mark the incident and outgoing light beams in free
space, as discussed in the text. Points B, C, D, E  mark the P-,
N-, N$_{ref}$- and P$_{ref}$-modes.} 
%
\vskip -0.3cm 
\label{f-Cheburashki}
\end{figure}
%

The size of the gap at $k_z = \pi/a$ is determined by the value of
$\varepsilon_c$ in Eq.(\ref{eps(z,w)}). In conventional PCs, the
contrast $\varepsilon_c$ comes from remote resonances (background
dielectric constant). In a gas with the mean density $\rho$,
substantial contrast appears only near an atomic resonance,
$\omega_T$ \cite{eberly}:
\begin{equation}\label{eps_res}
\frac{d}{a}\varepsilon_c \to \varepsilon_{\rm res}(\omega) =
\frac{8\pi \omega_T \mu^2 \rho}{\omega_T^2 - \omega^2 - 2 i
\omega_T \gamma},
\end{equation}
where $\mu$ is the dipole transition matrix element, and $\gamma$
accounts for losses. Below we imply that the periodic structure
shown on Fig.~\ref{f-Cheburashki}(b) depicts atoms in a 1D optical
lattice.
The calculations are done for the D2 line of Cs ($\mu =
4.48\,$a.u., $\w_T=11732\,$\cmm) with $\rho = 10^{13}~{\rm
cm}^{-3}$ \cite{10^13} in a lattice with $a=532\,$nm. Both
the Doppler and collisional widths at $T < 1$~K
are negligible compared to the radiative broadening $\gamma \simeq 33$~MHz \cite{Steck}.

{\textbf{\textit{Dynamics of coupled modes at the boundary.}}
%
{Inside the gas cloud, the dielectric contrast is a function of
the penetration depth: $\varepsilon_{\rm res}(x,\omega) =
\alpha(x) \varepsilon_{\rm res}(\omega)$. Here $\alpha(x)$ is the
density profile characterized by two scales: the total length of
the cloud $L$, and the length of the entrance and exit zones,
$L_*$. In these zones $\alpha(x)$ varies between 0 (vacuum) and 1
(saturated density).} Each value of $x$ can be assigned its own
EFS diagram. As $x$ grows, EFS gradually transforms from a single
circle $\sqrt{k_x^2+ k_z^2} = \w/c$ to a preiodic structure
similar to that shown in Fig.~\ref{f-Cheburashki}(c) by the thick
black line. The free-space mode shown by point A adiabatically
connects upon such gradual transformation with the P-mode
characterized by the same value of $k_z$ (point B). If the gas
density at the boundary changes slowly, then transfer of energy
from the P-mode both to the (wanted) N-mode (point C), and
(unwanted) reflected modes (points D and E) is suppressed.
%
%


We look for the field $E(x,z)$ inside the cloud in the form of a superposition of the N-
and P-modes. Coupling to the reflected modes is neglected on the
basis of higher adiabaticity, as explained below. Thus
\begin{equation}\label{Eofx-field}
E = \sum\limits_{Y = N,P} c_Y(x)\, | Y(x) \rangle e^{i \int_0^x
{\K}_Y(x_1)\,dx_1},
\end{equation}
where  ${\K}_{Y} = {\rm Re} [k_{Y}]$ for $Y=N,P$ is the real part
of $k_x=k_{N,P}$ found from Eq.(\ref{dispersion}). $| Y(x)
\rangle$ stands for the N- and P-eigenmodes of the type
(\ref{E-field}) calculated in the local basis: At each $x$, the amplitudes $C_m(x)$ of the partial waves are
related to the wave vector components $k_z, k_{Y}(x)$ as in
Eq.(\ref{C_m}) with $\varepsilon_c$ replaced by
$\alpha(x)\varepsilon_{\rm res}(\omega)$. For the $| N(x) \rangle$
mode we further denote $C_m \to N_m(x)$,  and for the $| P(x)
\rangle$ mode $C_m \to P_m(x)$. The damping enters through ${\rm
Im} [k_{N,P}(x)]$, and leads to decline of the amplitudes $c_N$
and $c_P$ as $x$ grows. Each eigenmode is normalized as $\langle Y
| Y \rangle \equiv \sum_m | Y_m|^2 = 4\pi \omega/c^2 {\K}_{Y}$
\cite{footnote-biOrthogonal}. This normalization corresponds to a
unit energy flow across the plane $x = const$ for zero damping.

To find the amplitudes $c_N(x)$ and $c_P(x)$ across the sample, we substitute
Eq.(\ref{Eofx-field}) into the wave equation with
$\varepsilon(x,z,\omega) = \alpha(x) \varepsilon(z,\omega)$, and
apply the slow envelope approximation \cite{eberly}, assuming that
the $x$-derivatives of $c_N(x)$, $c_P(x)$, $N_m(x)$, $P_m(x)$ are
all small compared to $\K_N, \K_P$. Using the fact that  the
vectors $|N\rangle, |P\rangle$ are the solutions of the wave
equation at a fixed $x$, we obtain:
%
\begin{eqnarray}
& \biggl[& 2{\K}_N c'_N | N \rangle +
       2{\K}_N c_N | N' \rangle +
       \left( {\K}'_N + i({\K}_N^2 - k_N^2) \right) c_N | N \rangle
\biggl]   \nonumber\\
&\times & e^{i \int_0^x dx_1 {\K}_{N}(x_1)} + \biggl[ 2{\K}_P c_P'
| P \rangle + 2{\K}_P c_P | P' \rangle  \biggr.
\nonumber \\
&\biggl. + &\left( {\K}'_P + i({\K}_P^2 - k_P^2) \right) c_P | P
\rangle \biggr] e^{i\int_0^x dx_1 {\K}_{P}(x_1) } = 0
\label{cN-cP-one_equation}
\end{eqnarray}
where `` $'$ '' stands for $x$-derivative. Multiplying
(\ref{cN-cP-one_equation}) consecutively by $\langle N |$ and
$\langle P |$ we find \cite{<N|N> etc}:
\begin{eqnarray}\label{cN-cP-system}
c'_N &=& \xi^* {\K}_P \, c_P  \exp\left[i \int_0^x
({\K}_{P}-{\K}_{N}) dx_1 \right] \,-  i \eta_N {\K}_N \, c_N
\\%
c'_P &=& -\xi {\K}_N \, c_N \exp\left[-i \int_0^x
({\K}_{P}-{\K}_{N}) dx_1\right] \, - i \eta_P {\K}_P \, c_P
~,\nonumber
\end{eqnarray}
where (note that ${\rm Im} \left[ \eta_Y {\K}_{Y} \right] = - {\rm Im} \left[  k_Y \right]$)
\begin{equation}
\xi = \frac{c^2}{4\pi\omega}  \langle P | N' \rangle, ~~~ \eta_Y =
\frac{c^2}{4\pi\omega} {\rm Im}\left[ \langle Y | Y' \rangle
\right] + \frac{1}{2}\left( 1 - \frac{k_Y^2}{{\K}_Y^2} \right).
\end{equation}
%
\begin{figure}
\includegraphics[width=0.99\columnwidth]{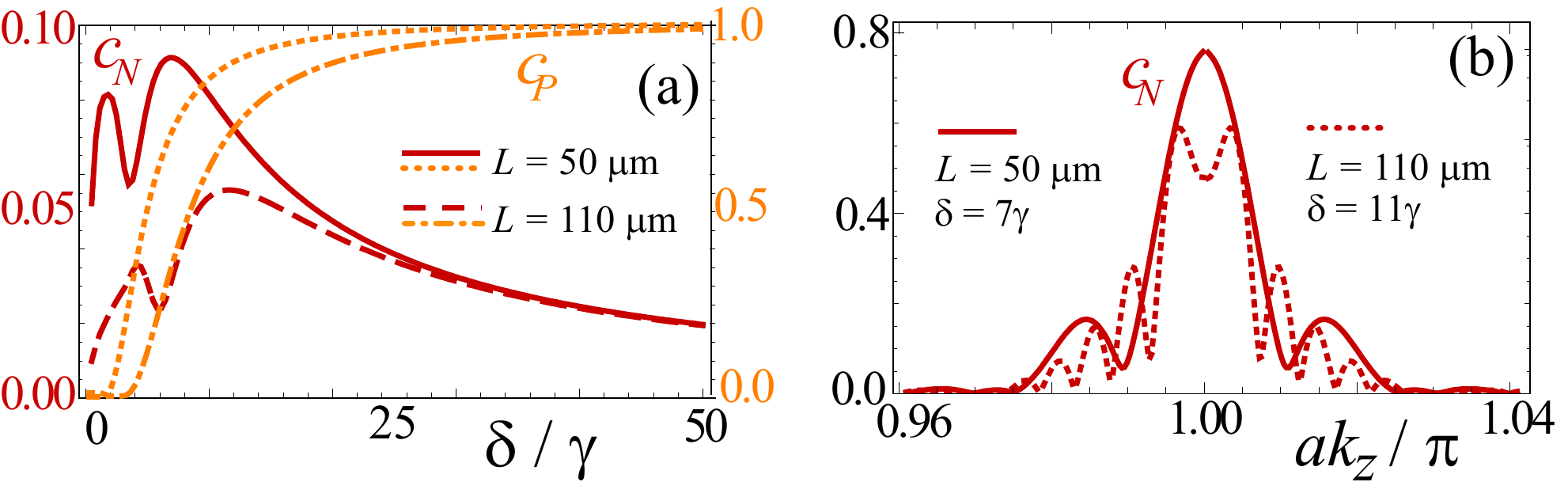}
\vskip -0.1cm 
\caption{(Color online) (a) Amplitudes $|c_P(L)|$ (gold, right
scale) and $|c_N(L)|$ (red, left scale) in dependance on $L$ and
$\delta$ with $L_{*} = 20~\mu$m. (b) $|c_N(L)|$ at $x=L$ as a
function $k_z=\w/c\sin\phi$ for two sets of parameters.}
\vskip -0.2cm 
\label{f-CN(L)CN(phi)}
\end{figure}
%
%

Eqs.(\ref{cN-cP-system}) resemble the equations describing
two-state quantum dynamics with the time derivative replaced by
$x$-derivative. We draw an analogy with the quantum adiabatic
theorem {\cite{Amin-Lidar}} by noticing that coupling between the
modes averages out if $|\xi {\K}_{P,N}| \ll |\K_{P}-\K_{N}|$ at
all $x$. When damping is negligible, $\eta_N \approx \eta_P
\approx 0$, the analogy can be further developed by noticing that
$ \langle P | N' \rangle$ is estimated as $\pi
\varepsilon_{\rm res}\, \omega /(2  v L_* \,  (k_P^2-k_N^2)) $}
with $v = 1/\left|\sum_{mn} P^*_m N_n \right|$
\cite{footnote-AdiabaticTheorem}. Using this estimate, we obtain
the adiabaticity condition as
\begin{equation}
\Omega_A \equiv L_{*} (k_P-k_N)^2 \,\frac{16 \,v}{\w
\varepsilon_{\rm res}} \gg 1~, \label{adiabaticity-last}
\end{equation}
$\Omega_A$ is the adiabaticity parameter. The transfer between the
modes only takes place if their wave vectors are sufficiently
close. The latter observation allowed us to neglect the reflected
modes in Eq.(\ref{Eofx-field}): At the values of $L_{*}$ such that
the evolution of coupled P- and N-modes corresponding to points B,
C in Fig.~\ref{f-Cheburashki}(c) is barely non-adiabatic, coupling
to the reflected modes corresponding to the points D, E can be
neglected due to larger $|k_P-k_{P_{ref},N_{ref}}|$. This is
different from conventional PCs, where reflection at the boundary
is always present.
%

{\textbf{\textit{Numerical results and discussion}}. For light
entering the sample, equations (\ref{cN-cP-system}) are solved
numerically with the boundary conditions $c_N(0)=0$, $c_P(0)=1$.
The amplitude $c_N(L)$ at the exit depends on four key parameters.
On one hand, $L_*$, $k_z = (\omega/c) \sin\phi$ and the detuning
$\delta=\w-\w_T$  determine the adiabaticity parameter $\W_A$. On
the other hand, $\delta$ and $L$ determine losses due to absorption. The smaller is $\delta$,
the larger is the (wanted) dielectric contrast and (unwanted)
absorption. In Fig.\ref{f-CN(L)CN(phi)}(a) we plot the amplitudes
at the exit, $|c_N(L)|$, $|c_P(L)|$, as functions of $\delta$ for
$L=50$ and 110~$\mu$m, with $L_{*} = 20~\mu$m, $\rho=10^{13}~{\rm
cm}^{-3}$, $k_z = 0.98 \pi/a$, and $\alpha(x)=1$ inside the cloud
and changing according to $\sin^2 (\pi x/2L_{*})$ law at the
boundaries. The amplitude $c_N(L)$ reaches values $0.05 \div 0.1$
in the frequency window as large as $20 \g$, i.e. over 2.5 GHz,
slowly declining at larger detunings. For strong absorption
($\delta <~ 2\g$ for our geometry), coupling between the modes is
only effective at the entrance: The P-mode is completely absorbed
at $x\lessapprox L_*$, see the dynamics of $c_{P,N}(x)$ in   
Fig.\ref{f-Zaichik-strong-absorption}(c). The N-mode survives,
experiencing much lower absorption, due to reasons
discussed below. 
In the regime of small absorption ($\delta >~ 20\g$) $c_P(L)\simeq
1$, and $c_{P,N}(L)$ do not depend on $L$. We observed that at
higher densities the energy transfer dynamics strongly resembles
that in a Landau-Zener transition \cite{Raizen}}. At intermediate
absorption, the values $c_P(L)$ and $c_N(L)$ are comparable.
Fig.\ref{f-CN(L)CN(phi)}(b) shows the angular dependence of
$|c_N(L)|$ and $|c_P(L)|$ calculated for the same density profiles
as in Fig.\ref{f-CN(L)CN(phi)}(a). The value $k_z=0.98 \pi/a$,
used in the rest of our calculations for illustrative purposes, is
at the edge of the window of allowable angles. Closer to the Bragg
angle, $|c_N(L)|$ can be as high as 0.8.

\begin{figure}[b]
\centering
\includegraphics[width=\columnwidth]{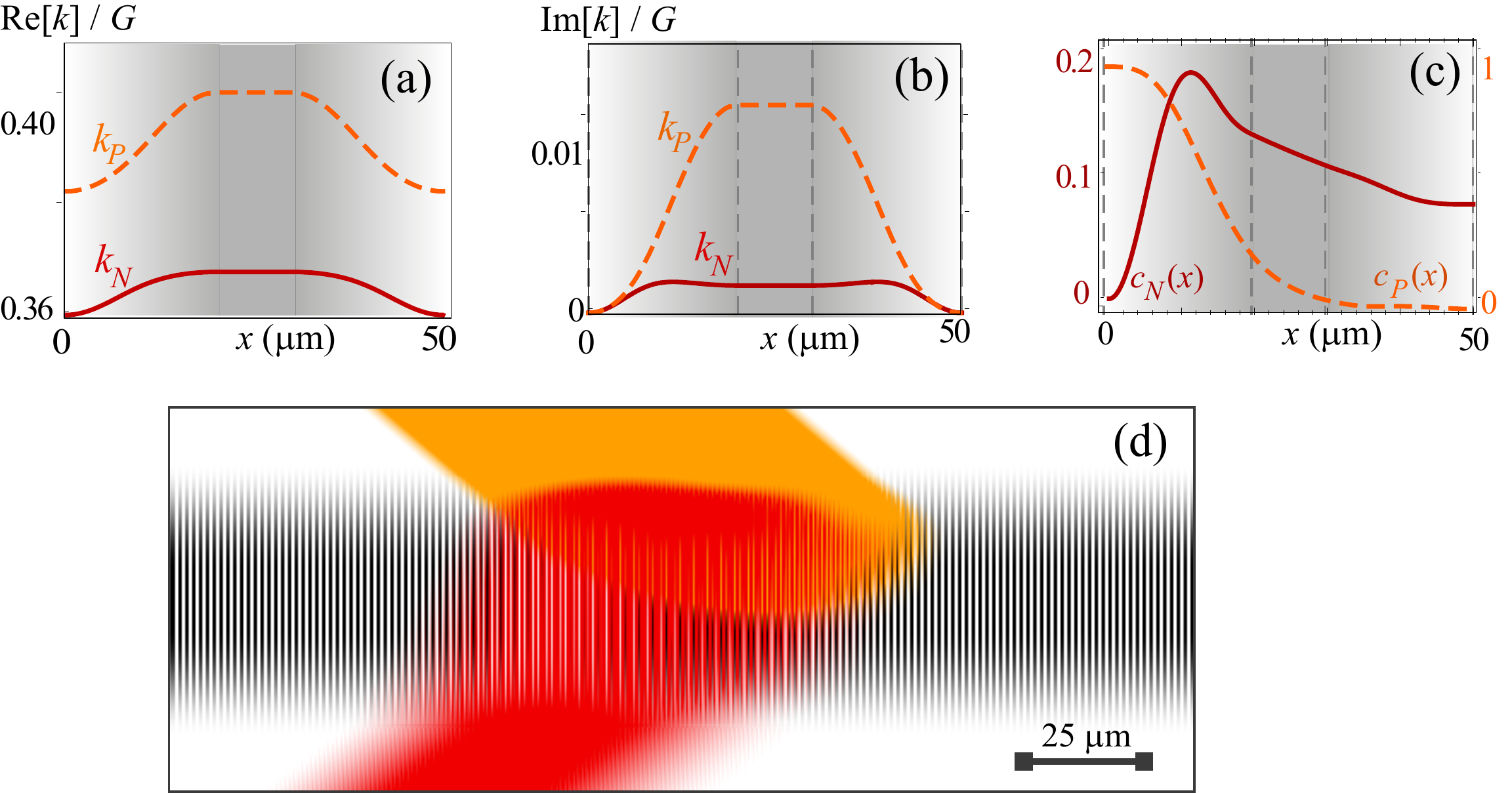}
\caption{(Color online) Strong absorption ($\delta = 2 \gamma$).
Spatial dependence of the real (a) and imaginary (b) parts of
$k_N$ and $k_P$, and of $c_N$ and $c_P$ (c). Dashed vertical lines
mark the entrance and exit zones. (d) Beam propagation through the
cloud. Gold shows P-beam, red -- N-beam.}
\vskip -0.2cm 
 \label{f-Zaichik-strong-absorption}
\end{figure}

Figure~\ref{f-Zaichik-strong-absorption} illustrates propagation
of a Gaussian beam  with the central wave vector $k_z = 0.98
\pi/a$, $\delta = 2 \gamma$, through a cloud with $L_{*} =
20$~$\mu$m, $L=50$ \mum.
The intensities of the two modes in Panel (d) are calculated by
expanding the initial Gaussian beam into plain waves, each with
its own $k_x,k_z$. For each of them we write the field as in
Eqs.(\ref{C_m},\ref{Eofx-field}), and propagate it according to
Eqs.(\ref{cN-cP-system}). Then we combine the waves to retrieve
the overall field.
The resulting angles of propagation of the P- and N-beams,
$\phi_N$ and $\phi_P$, correspond to the curvature of the  EFS at
points B,C in Fig.~\ref{f-Cheburashki}(c).

According to Panels (a) and (c), transfer between the modes is
only efficient for $x \leq 10~\mu$m, where the distance between
$\K_N$ and $\K_P$ is minimal and dynamics have non-adiabatic features, cf. Eq.(\ref{adiabaticity-last}). %
The P-mode is completely absorbed inside the cloud, and only the
N-mode is present at the exit. As the gas density at the exit of
the cloud decreases, all the amplitudes $N_m$ except $N_{-1}$,
depicted by point C$'$ in Fig.~\ref{f-Cheburashki}(c), vanish. Due
to partial adiabaticity of the exit dynamics, C$'$ connects with A$'$ on the green circle, and the N-mode leaves the cloud in such a way that its NR-like propagation is preserved.

Panel (b) of Fig.\ref{f-Zaichik-strong-absorption} illustrates a
key ingredient of our scheme. As the gas density increases, ${\rm
Im}[k_{P}]$ increases as well, but ${\rm Im}[k_N]$ quickly reaches
maximum and stabilizes. The closer $k_z$ to $\pi/a$ or the higher
gas density, the smaller  ${\rm Im}[k_{N}]$. From Eq.(\ref{C_m}),
near $k_z\simeq \pi/a$ the partial wave amplitudes $P_m, N_m$ with
$m=0,-1$ are large, and all others are small. For the N-mode
$N_0\simeq -N_{-1}$, whereas $P_0 \simeq P_{-1}$
\cite{footnote-AlternatingSigns}. Stabilization of the asymmetric
N-mode can be attributed to destructive interference: the field
maxima are at the $z$ values with no lattice atoms. For the
symmetric P-mode, the field maxima coincide with the lattice
density maxima, and absorption is high. This phenomenon can be
described \cite{Berry-grating} in the language similar to that of
many interference phenomena in quantum mechanics
\cite{QM-stabilization}. The absence of absorption in the N-mode
allows  choosing small detunings from the resonance, thus
weakening the requirement for the gas density.

Fig.\ref{f-Zaichik-weak-absorption} shows propagation of the same
Gaussian beam as in Fig.\ref{f-Zaichik-strong-absorption} for
intermediate absorption, $\d=3\g$ and $7\g$. By varying the
detuning, one can control the P-mode exit intensity while keeping
the N-mode intact. For a large detuning and low absorption (Panels
(c), (d)), non-adiabaticity of the exit with $L_*=20~\mu$m begins
to play a role: The amplitude $c_N$ grows both at the entrance and
at the exit from the cloud. Counterintuitively, the N-mode is
generated in the regions of low, rather than high, gas density --
i.e. where adiabaticity is low.

\begin{figure}
\centering
\includegraphics[width=\columnwidth]{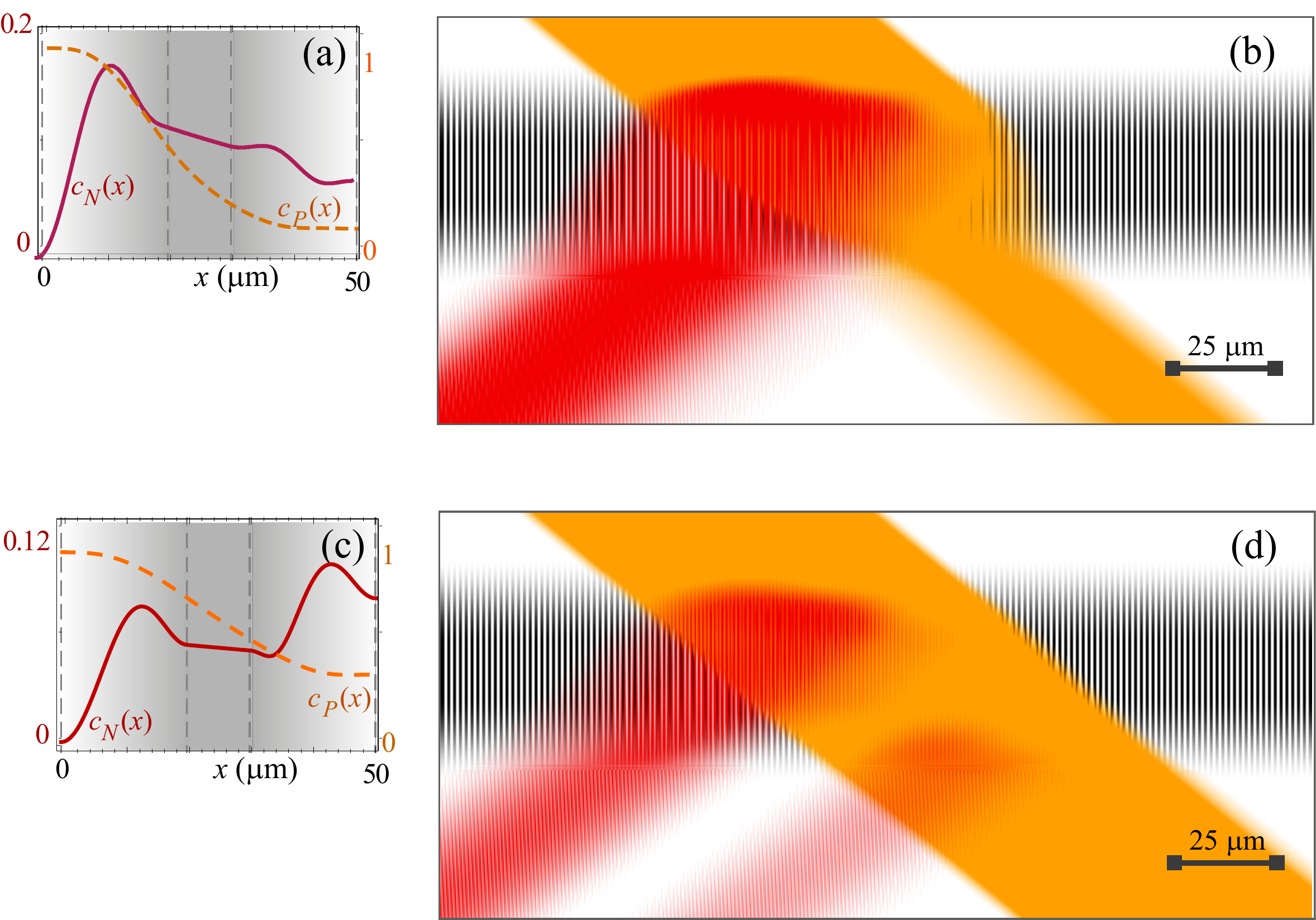}
\caption{(Color online) Same as
in Fig.\ref{f-Zaichik-strong-absorption} (c,d), but for
intermediate and weak absorption: $\delta = 3 \gamma$ (a,b), and
$\delta = 7 \gamma$ (c,d).}
\vskip -0.2cm 
\label{f-Zaichik-weak-absorption}
\end{figure}

\textbf{\textit{In conclusion,}} our scheme realizes NR-like light
propagation in a cold gas at the experimentally achieved density
of $10^{13}~{\rm cm}^{-3}$. This density estimate is three orders
of magnitude lower than in the chirality-based proposals
\cite{Chirality}, and five orders of magnitude lower than in
magnetic resonance-based proposals \cite{Turki}. The advantage is
due to the PC-like structure being induced via strong
electric-dipole couplings in atoms, as compared to weak
magnetic-dipole couplings required in previous schemes.
Stabilization of the N-mode against absorption further weakens the
density requirements. Finally, due to the effect of the grating, a
very weak contrast of $\varepsilon$ (in our calculations,
$\varepsilon_c\sim 10^{-2}$) is sufficient for strong modification
of light propagation. At the same time, bandwidth of the NR window
is $\sim20$ times higher than {the frequency window in EIT-based
bandgap structures} \cite{Lukin-LaRocca}.


Implementation of our scheme in higher-dimensional lattices with
true negative refraction is straightforward. In conventional PCs,
high dielectric contrast is required to avoid the unwanted
positively refracted wave \cite{Notomi}. The present scheme can be
employed even with small $\varepsilon_c$, since the birefrigence
is suppressed via absorption of the P-mode. Vaguely defined
boundaries make the dynamics of light conceptually different from
that at a conventional interface. However, with reasonable length
parameters one can transfer noticeable fraction of light into the
N-mode while fully controlling intensity of the P-mode at the
exit. Intensity of the negatively refracted light will be higher
for higher gas densities, smaller absorption, and angles closer to
the Bragg angle.


{\textbf{\textit{Acknowledgements}}. The authors dedicate this
work to the memory of Moshe Shapiro. His insight stimulated our
work on negative refraction in gases.

We are pleased to thank 
M. Sukharev and K. Madison for consultations.


\begin{thebibliography}{99}

\bibitem{NR}
``Physics of negative refraction and negative index materials'', C. M. Krowne and Y. Zhang (eds.), Springer, 2007.


\bibitem{invisibility}
J. Pendry,
Physics, {\bf 2}, 95 (2009);
%
Y. Lai et al,
Phys. Rev. Lett., {\bf 102}, 253902 (2009).

\bibitem{lens}
V. G. Veselago, Sov. Phys. Uspekhi, {\bf 10}, 509 (1968);
%
J. B. Pendry,
Phys. Rev. Lett., {\bf 85}, 3966 (2000);
%
P. V. Parimi, W. T. Lu, P. Vodo, S. Sridhar,
Nature {\bf 426} 404 (2003).

\bibitem{corner}
See http://skullsinthestars.com/2009/05/19/what-does-negative-refraction-look-like/ and references therein.

\bibitem{superprism}
H. Kosaka et al.,
Phys. Rev. B {\bf 58}, R10096 (1998).

\bibitem{Turki}
M. $\ddot{\rm O}$. Oktel, $\ddot{\rm O}$. E. M$\ddot{\rm
u}$stecaplio$\breve{g}$lu,
Phys. Rev A, {\bf 70}, 053806 (2004).

\bibitem{Chirality}
J. K$\ddot{\rm a}$stel, M. Fleischhauer, S. F. Yelin,  R. L. Walsworth,
Phys. Rev. Lett., {\bf 99}, 073602 (2007);
%
D. E. Sikes, D. D. Yavuz,
Phys. Rev. A {\bf 82}, 011806(R) (2010).


\bibitem{Kravtsov}
V. E. Kravtsov, V. M. Agranovich, K. I. Grigorishin,
Phys. Rev. B, {\bf 44}, 4931 (1991).

\bibitem{Foteinopoulou}
E. Cubukcu, K. Aydin, E. Ozbay, S. Foteinopoulou, C.M. Soukoulis, %
Nature {\bf 423}, 604 (2003). For a review, see S. Foteinopoulou,
Physica B {\bf 407}, 4056 (2012) and references therein.

\bibitem{Notomi}
M. Notomi,
Phys. Rev. B {\bf 62},  10696 (2000);
%
S. Foteinopoulou, C. M. Soukoulis, %
Phys. Rev. B {\bf 72}, 165112 (2005).

\bibitem{EIT-exp}
G. Birkl, M. Gatzke, I. H. Deutsch, S. L. Rolston, W. D. Phillips, %
Phys. Rev. Lett. {\bf 75}, 2823 (1995);
%
I. H. Deutsch, R. J. C. Spreeuw, S. L. Rolston, W. D. Phillips, %
Phys. Rev. A {\bf 52}, 1394 (1995);
%
A. Schilke, C. Zimmermann, P. W. Courteille, W. Guerin %
Phys. Rev. Lett. {\bf 106}, 223903 (2011).

\bibitem{Lukin-LaRocca}
A. Andre and M. D. Lukin, %
Phys. Rev. Lett. {\bf 89}, 143602 (2002);
%
D. Petrosyan, %
Phys. Rev. A {\bf  76}, 053823 (2007);
M. Artoni and G.C. La Rocca, %
Phys. Rev. Lett. {\bf 96}, 073905 (2006);
%
Y. Zhang {\it et. al},
Op. Ex. {\bf 21}, 29338 (2013).
%

\bibitem{10^13}
C. Chin, A. J. Kerman, V. Vuleti$\acute{\rm c}$, S. Chu, Phys. Rev. Lett., {\bf 90}, 033201 (2003).

\bibitem{Russel}
P. St. J. Russel,
Appl. Phys. B, {\bf 39}, 231 (2986).

\bibitem{Chu-Tamir}
R. S. Chu and T. Tamir, %
Proc. IEEE, {\bf 119},  197 (1972);
%
R. S. Chu, J. A. Kong,
IEEE Trans. of Microwave Theory and Techniques, MTT-{\bf 25}, 18 (1977).

\bibitem{eberly}
L. Allen, J. H. Eberly, %
''Optical resonance and two-level atoms''%
(Wiley-Interscience, New York, 1975).

\bibitem{Steck}
Daniel A. Steck, Cesium D Line Data, available online at http://steck.us/alkalidata (revision 2.0.1, 2 May
2008).

\bibitem{footnote-biOrthogonal}
We adopt ``bra'' and ``ket'' notations of conventional quantum
mechanics. Alternatively, one can use the formalism of
biorthogonal states, see e.g.
%
N. Moiseyev, ``Non-Hermitian quantum mechanics'', (Cambrige
University Press, Cambrige, 2011)
%
and Refs.\cite{Fedorov-lattice, Atabek}.

\bibitem{Fedorov-lattice}
M. V. Fedorov, M. A. Efremov, V. P. Yakovlev, W. P. Schleich,
JETP {\bf 97}, 522 (2003).

\bibitem{Atabek}
O. Atabek and R. Lefebvre, %
J. Phys. Chem. A  {\bf 114}, 3031 (2010).


\bibitem{<N|N> etc}
We used the relations ${\rm Re} \left[ \langle Y | Y' \rangle
\right] = -2\pi\omega {\K}'_Y/c^2 {\K}_Y^2$ and $\langle N | P'
\rangle = \langle P' | N \rangle^* = -\langle P | N' \rangle^*$,
obtained by differentiating $\langle Y | Y \rangle = 4\pi
\omega/c^2 {\K}_{Y}$ and $\langle P | N \rangle = 0$ over $x$.

\bibitem{Amin-Lidar}
%
M. S. Sarandy and D. A. Lidar, %
Phys. Rev. A {\bf 71}, 012331 (2005);
%
A. Fleischer and N. Moiseyev, %
Phys. Rev. A {\bf 72}, 032103 (2005);
%
M. H. S. Amin, %
Phys. Rev. Lett. {\bf 102}, 220401 (2009).

\bibitem{footnote-AdiabaticTheorem}
We introduce a local-basis wave operator $\hat L(x) =
\partial^2/\partial z^2 + \ve(z,\w)\w^2/c^2$. At a given $x$,
$\hat L(x) |N(x)\rangle = k_N^2(x) |N(x)\rangle$; similar for
$|P\rangle$ and $k_P$. We diffierentiate the last equation over
$x$, multiply the resut by $\langle P |$ and use the wave equation
for the P-mode $\langle P | \hat L = k_P^2 \langle P |$. We then
replace $\langle P | \hat L' | N \rangle$ by $\a' \langle P |
\ve(\w,z) | N \rangle = \a' \ve_c \sum_{nm} P^*_m N_n$. For the
density profile used in the calculations, we replace $\alpha'$ by
$\pi/2L_{*}$, and notice that in the regions of low adiabaticity
$k_P\simeq k_N$, and $|k_{P,N} / (k_P+ k_N)| \simeq 1/2$.

\bibitem{Raizen}
Q. Niu and M. G. Raizen, %
Phys. Rev. Lett. {\bf 80}, 3491 (1998).

\bibitem{footnote-AlternatingSigns}
The sign of the amplitude $C_m$ depends on that of the denominator
in Eq.(\ref{C_m}). The latter can be visualized using
Fig.~\ref{f-Cheburashki}(c), and depends on whether the point on
EFS, which corresponds to the $m$-th partial wave, is inside or
outside the green dashed circle with the radius $\w/c$. For the
N-mode, point C representing the 0$^{th}$ partial wave is inside
the circle, and point C$'$ representing the -1$^{st}$ wave is
outside. Thus the signs of $N_0$ and $N_{-1}$ are different. For
the P-mode, both 0$^{th}$ and -1$^{st}$ partial waves are
represented by points outside the green circle, and the signs of
$P_0$ and $P_{-1}$ coinside.

\bibitem{Berry-grating}
See M. V. Berry, D. H. J. O'Dell,
J. Phys. A: Math. Gen., {\bf 31}, 2093 (1998). To compare with our
case, one must assume complex $\sigma$ in Eq.(8), and replace in
Eq.(12) $\sigma/2$ by $\sigma$ to account for $\delta(x)$-like
instead of $\cos$-like perturbation.


\bibitem{QM-stabilization}
The stabilization phenomenon is described in
Ref.\cite{Berry-grating} in terms of the amplitudes of 0$^{th}$
and -1$^{st}$ partial waves coupled via the lattice.
Compare, e.g., with strong-field interference stabilization in %
M. V. Fedorov, N. P. Poluektov, A. M. Popov, {\it et. al.}, 
IEEE J. Sel. Papers on Quantum Electronics {\bf 18}, 42 (2002);
%
formation of dark states via interference and EIT in %
M. Shapiro and P. Brumer, %
``Quantum Control of Molecular Processes'' %
(Whiley-VCH, Weinheim, 2012) %
and %
M. Fleischhauer, A. Imamoglu, J. P. Marangos,
Rev. Mod. Phys. {\bf 77}, 633 (2005),
%
as well as stabilization of atoms passing through an optical
lattice in \cite{Fedorov-lattice} and resonances of dissociating
molecules in \cite{Atabek}.




%
%
%

%
%





%
%


%


\end{thebibliography}
\end{document}